\documentclass[12pt]{article}
\usepackage[left=2cm,right=2cm]{geometry}
\usepackage{graphicx}

\title{Solution to Faddeev equations with two-body experimental amplitudes as input and application to $J^P=1/2^+$, $S=0$ baryon resonances}
\author{A. Mart\'inez Torres\textrm{$^{1,~2}$}, K. P. Khemchandani\textrm{$^{2}$}, and E. Oset\textrm{$^1$}
\footnote{amartine@ific.uv.es, kanchan@teor.fis.uc.pt, oset@ific.uv.es}}

\date{}
\begin{document}
\maketitle
\begin{center}
\textrm{$^1$} Departamento de F\'{\i}sica Te\'orica and IFIC,
Centro Mixto Universidad de Valencia-CSIC, Institutos de
Investigaci\'on de Paterna, Aptdo. 22085, 46071 Valencia, Spain.\\
\vspace{0.5cm}
\textrm{$^2$} Centro de F\' isica Te\'orica, Departamento de F\'isica, Universidade de Coimbra, P-3004-516 Coimbra, Portugal.
\end{center}
\maketitle
\begin{abstract}
We solve the Faddeev equations for the two meson-one baryon system $\pi\pi N$ and coupled channels using  the experimental two-body $t$-matrices for the $\pi N$ interaction as input and unitary chiral dynamics to describe the interaction between the rest of coupled channels. In addition to the $N^*(1710)$ obtained before with the $\pi\pi N$ channel, we obtain, for $J^\pi=1/2^+$ and total isospin of the three-body system $I=1/2$, a resonance peak whose mass is around 2080 MeV and width of 54 MeV, while for $I=3/2$ we find a  peak around 2126 MeV and 42 MeV of width. These two resonances can be identified with the $N^* (2100)$ and the $\Delta (1910)$, respectively. We obtain another peak in the isospin $1/2$ configuration, around $1920$ MeV which can be interpreted as a resonance in the $N a_0(980)$ and $N f_0(980)$ systems.
\end{abstract}
\section{Introduction}
 Recent developments around three-body systems with two mesons and one baryon
using chiral dynamics have brought new light into the nature of the $J^P=1/2^+$
baryonic resonances. The study of such systems with strangeness $S=-1$ produced
resonant states which could be identified with the existing low lying baryonic
$J^P=1/2^+$ resonances, two $\Lambda$ and four $\Sigma$ states   
\cite{MartinezTorres:2007sr,MartinezTorres:2008is,Khemchandani:2008zz}. Similarly, in the case of
the $S=0$ sector the $N^*(1710)$ appears neatly as a resonance of the $\pi \pi N$
system, with or without including its coupled channels  within SU(3)
\cite{Khemchandani:2008rk}.  Developments along the same direction produced a
resonant state of $\phi K \bar{K}$ \cite{MartinezTorres:2008gy} which could be 
identified with the X(2175) resonance reported at BABAR 
\cite{Aubert:2006bu,Aubert:2007ur} and later on at BES \cite{:2007yt}. The study
of the three-body systems was done using Faddeev equations (FE) in the coupled channel 
approach. While most conventional studies of three-body systems 
use potentials in coordinate space, usually separable potentials to make the
solution of the FE feasible, the approach of 
\cite{MartinezTorres:2007sr,MartinezTorres:2008is} used two particle amplitudes
generated within the unitary chiral approach in momentum space. Yet, the most 
novel finding in these works was the realization that, for s-waves and in the 
SU(3) limit, there was an exact
cancellation between the off shell part of the
two-body amplitudes and the three-body forces generated by the same chiral
Lagrangians. To be more precise, the on shell amplitude
means that the s-wave amplitude is calculated as a function of the
Mandelstam variable $s$ imposing $q^2=m^2$ for the external momenta of the two
body amplitudes. When these lines are inside the Faddeev diagrams where some
line can be off shell, the
full amplitude is separated into this ``on shell" part plus and ``off shell" part
which goes as $q^2-m^2$ for mesons and $q^0-E(q)$ for baryons and vanishes when
the external lines are on shell.  This off shell part contains an inverse
particle propagator and cancels one particle propagator rendering a Faddeev
diagram with two two-body t-matrices into a three-body contact term, which has
the same topology as genuine three-body interactions that stem from the 
chiral Lagrangians and cancel them exactly. As a consequence, one needs only
the on shell two-body t-matrices and can ignore these three-body forces. This
finding is novel for such studies and simplifies the work technically, although
not much, since loops involve a changing s-variable, and consequently the
s-dependent t-matrices must be inserted into the loop functions. This makes
this approach different and technically more involved than the study of the two
body interaction, where using arguments of the N/D method one can factorize on
shell amplitudes outside the loop functions which involve only two hadron
propagators \cite{Oller:1998zr,Oller:2000fj}. The strongest value of that
finding in the three-body problem is 
that the results do not depend upon the off shell extrapolations of the 
amplitudes which is a source of uncertainty in the three-body calculations 
that rely upon a potential. Indeed, it is well known that given a certain
physical amplitude, on shell by nature, one has an infinite number of 
potentials that give this amplitude upon solving the Schr\"odinger equation. The
differences between the different potentials will only show in the off shell
extrapolation of the amplitudes. However, this information enters the solution
of the Faddeev equations and, hence, different potentials leading to the same on
shell amplitude will provide different results upon solution of the Faddeev
equations. 

    The problem stated above is most probably the main reason why recent works
dealing with the $\bar{K} NN$ system lead to quite different results in the
binding and the width. In this sense, we find a series of works based on Faddeev
equations which lead to
relatively large binding, of the order of $50-70$ MeV 
\cite{Shevchenko:2006xy,Shevchenko:2007zz,Yamazaki:2007cs,Ikeda:2007nz}, while
other works based on variational methods lead to smaller bindings of the order
of 20-30 MeV \cite{Dote:2007rk,Dote:2008in,Dote:2008hw}. The widths also vary
from $50-100$ MeV. 

   The arbitrariness of the off shell amplitude is also well known in field
theory, where the implementation of unitary transformations of the fields in the
Lagrangian maintains the same on shell amplitudes but changes their off shell
extrapolation.  In this sense it is interesting to note that, 
although the off shell versus three-body cancellation 
discussed here is not explicitly shown in other three-body works using also
chiral dynamics \cite{Epelbaum:2002ji,Bernard:2007sp}, the approach is invariant
upon these transformations, indicating that the mentioned cancellations
apparently occur in the full calculation \cite{hanhart}. A similar independence
on the off shell extrapolation has been shown in different reactions like  
the $\pi N \to \pi \pi N$ reaction \cite{Hanhart:2007mu} and the study of the
interacting two pion exchange in the $NN$ interaction \cite{Oset:2000gn}.
However, the explicit realization of the off shell versus three-body forces
indicates that one can neglect the three-body forces from the beginning, 
certainly simplifying the approach, and use only the two-body on shell 
amplitudes.  Even more, these on shell amplitudes can be obtained from
experiment and one can omit having to do a theory for the two-body
interaction. There is a small caveat there, since sometimes in the loops one
will need ``on shell" amplitudes below threshold. This looks like a
contradiction, but we made it clear the meaning on the on shell amplitude needed
in the Faddeev equations, which is the one where $q^2=m^2$ for the external 
momenta. Provided one has a suitable parameterization of the amplitude, the
extrapolation below threshold fulfilling this condition is not a difficult task
to accomplish. In many cases, like in the present one that we shall discuss
here, one needs the information
well within the physical range and the extrapolation is not even needed. This
picture presented here is rather novel and the purpose of the present paper is
to show how it works and how it can help whenever the theoretical models are not
accurate enough. 

   With the perspective given above we shall tackle here the investigation of
three-body systems with two mesons and a baryon with strangeness $S=0$.  The
problem was already discussed in \cite{Khemchandani:2008rk}, where the 
$N^*(1710)$ was found as a resonant state of $\pi \pi N$. It was also found
there that the implementation of other coupled channels barely changed the
results obtained with the base of the $\pi \pi N$ states alone. Yet, there
are other $J^P= 1/2^+$ states, like the $N^*(2100)$ and the $\Delta(1910)$, which
do not appear with that base and the use of the amplitudes obtained with the
lowest order chiral lagrangians. From the work of \cite{Inoue:2001ip} we know
that the chiral unitary approach using the lowest order chiral Lagrangian
provides a fair amplitude up to $\sqrt{s}= 1600~ MeV$ but fails beyond this
energy. For instance, the $N^*(1650)$ does not appear in the approach. As a
consequence, any three-body states which would choose to cluster a $\pi N$
subsystem into this resonance would not be obtained in the approach of 
\cite{Khemchandani:2008rk}. In the present work we shall give the step to
use experimental  $\pi N$ amplitudes and will show that in this case we
reproduce the  $N^*(1710)$ resonance without practically any modification with
respect to \cite{Khemchandani:2008rk}, but the use of a more realistic $\pi N$
interaction at higher energies leads also to the generation of the 
 $N^*(2100)$ and the $\Delta(1910)$ resonances as three-body systems of two
 mesons and one baryon in coupled channels. 
 
\section{Formalism and Results}
We follow the method developed in  \cite{MartinezTorres:2007sr,MartinezTorres:2008is,Khemchandani:2008zz,Khemchandani:2008rk,MartinezTorres:2008gy} to calculate the three-body $T$-matrix and search for resonances. In  \cite{MartinezTorres:2007sr,MartinezTorres:2008is,Khemchandani:2008zz,Khemchandani:2008rk,MartinezTorres:2008gy} a coupled channel Bethe-Salpeter equation is solved to calculate the required two-body $t$-matrices with the potentials obtained from chiral Lagrangians. These $t$-matrices, which contain the information of the two-body resonances, are then used as an input to solve the Faddeev equations in a coupled channel approach. The Faddeev equations,

\begin{equation}
T^{i}=t^i\delta^3(\vec{k}^{\,\prime}_i-\vec{k}_i) +
t^i g^{ij}T^j + t^i g^{ik}T^k , \quad\textrm {$i\neq j\neq k=1, 2, 3$}\label{Fa1}
\end{equation}
in our formalism, have been reformulated to
\begin{equation}
T^i\equiv t^i\delta^3(\vec{k}^{\,\prime}_i-\vec{k}_i) +T^{ij}_{R}+T^{ik}_{R},\nonumber
\end{equation}
where $T^{ij}_{R}$ satisfy the equations
\begin{equation}
T_R^{ij}=t^ig^{ij}t^j+t^i[G^{ijk}T_R^{jk}+G^{iji}T_R^{ji}].\label{Fa}
\end{equation}

Eqs. (\ref{Fa}) are  matrix equations since we solve them for coupled channels.  In Eqs. (\ref{Fa}) , $t^n$'s are the two-body $t$-matrices for the interaction of the pair of particles $(ml)$ , with $n \neq m \neq l$ ($n$ is the index of the spectator particle) and the elements of the $g^{ij}$ matrix  can be written in a general form as

\begin{equation}
g^{ij} (\vec{k_i}^\prime, \vec{k_j}) = \Biggr( \prod_{r=1}^D 
\frac{N_r}{2 E_r} \Biggr) \frac{1}{\sqrt{s} - E_i (\vec{k_i}^\prime) - E_l(\vec{k_i}^\prime + \vec{k_j}) - E_j(\vec{k_j})}, \vspace{1cm}  l \ne i \ne j=1,2,3 \label{gij}
\end{equation}
where $D$ is the number of particles propagating between two $t$-matrices. Following the normalization of \cite{mandlshaw}, $N_r = 1$ for a meson and $N_r = 2 M_r$ for a baryon with $M_r$ being the mass of the baryon and $\vec{k_i}^\prime (\vec{k_j})$ represent the momentum of the $i$th ($j$th) particle in the final (initial) state (in the global center of mass system).

The $G^{ijk}$ matrix in Eqs. (\ref{Fa}) is defined in terms of the product of two matrices as
\begin{equation}
G^{ijk} = \int\frac{d^3 k^{\prime\prime}}{(2\pi)^3} \tilde{g}^{ij}(s_{lm}, \vec{k^{\prime\prime}}) F^{ijk} (\vec{k^{\prime\prime}},\vec{k_j^{\prime}},\vec{k_k},s_{l^\prime k^\prime}^{k^{\prime\prime}}) \quad i\neq j,\, j\neq k=1, 2, 3 \label{eq:G}
\end{equation}
with $l \ne m \ne i, l^\prime \ne k^\prime \ne j$, where the elements of the $\tilde{g}^{ij}$ matrix are 
given by

\begin{equation}
\tilde{g}^{ij}=  \frac{N_l}{2E_l(\vec{k}^{\prime\prime})} \frac{N_m}{2E_m (\vec{k}^{\prime\prime})}
\frac{1}{\sqrt{s_{lm}}-E_l(\vec{k}^{\prime\prime})-E_m(\vec{k}^{\prime\prime})
+i\epsilon}\label{gtilde}
\end{equation}
and the $F^{ijk}$ matrix is defined as
\begin{equation}
F^{ijk} (\vec{k^{\prime\prime}},\vec{k_j^{\prime}},\vec{k_i},s_{l^\prime k^\prime}^{k^{\prime\prime}}) = t^j( s_{l^\prime k^\prime}^{k^{\prime\prime}} ) \times g^{jk} (\vec{k^{\prime\prime}}, \vec{k}_k) \times [g^{jk} (\vec{k}_j^\prime, \vec{k}_k)]^{-1} \times [ t^j  (s_{l^\prime k^\prime})]^{-1}.\label{eq:f}
\end{equation}

In Eq. (\ref{gtilde}), $E_l(\vec{k}^{\prime\prime}) = \sqrt{(\vec{k}^{\prime\prime})^2 + m_l^2}$ and 
$\sqrt{s_{lm}}$ is the invariant mass of the $(lm)$ pair (with $i \neq l \neq m$). We solve Eqs. (\ref{Fa}) as a function of the total energy of the system, $\sqrt{s}$, and the invariant mass of the particles $2$ and $3$, $\sqrt{s_{23}}$. The rest of the kinematical variables, i.e., the other $s_{ij}$'s, energies and momenta of the three particles in different reference frames, are defined in terms of these two variables as shown in detail in \cite{Khemchandani:2008rk,MartinezTorres:2007sr}.
In Eq. (\ref{eq:f}) the $k^{\prime\prime}$ superindex on $s_{l^\prime k^\prime}^{k^{\prime\prime}}$ indicates that the definition of this invariant mass depends on the loop variable.

Eq. (\ref{eq:f}) is derived in such a way that diagrams up to three $t$ matrices are calculated exactly by multiplying the contribution of diagrams with two $t$ matrices by the loop function of Eq. (\ref{eq:G}) and one more $t$ matrix. This same iteration procedure is used to go from diagrams containing  three to four $t$ matrices and so on. The procedure was proved to be numerically accurate in \cite{Khemchandani:2008rk}. As one can see, the formalism has been developed in such a way that all the loop variable dependence is contained in the $G^{lmn}$-functions and thus  Eqs. (\ref{Fa}) are an algebraic set of equations.

We solve Eq. (\ref{Fa}) for the $\pi\pi N$ system and coupled channels. The present work benefits from the previous study of the $\pi\pi N$ system and coupled channels \cite{Khemchandani:2008rk}, where the dynamical generation of the  $N^*(1710)$ was found. This state was found when the total isospin $1/2$ three-body $T$-matrix was evaluated by adding a nucleon to the two pions interacting in isospin zero in the energy range of the $\sigma$-resonance. The $N^*(1710)$ was thus interpreted as a resonance in the $\pi\pi N$ system, where the two pions rearrange to form the $\sigma$-resonance. The total energy range studied in \cite{Khemchandani:2008rk} corresponded to a variation of the invariant masses of the $\pi N$ pairs up to $\sim$ 1550 MeV. The calculations in \cite{Khemchandani:2008rk} were limited to this energy range because the input $\pi N$ $t$-matrix used in that work was taken from \cite{Inoue:2001ip} which reproduces the $\pi N$ scattering data well up to about 1600 MeV.

\begin{figure}[ht]
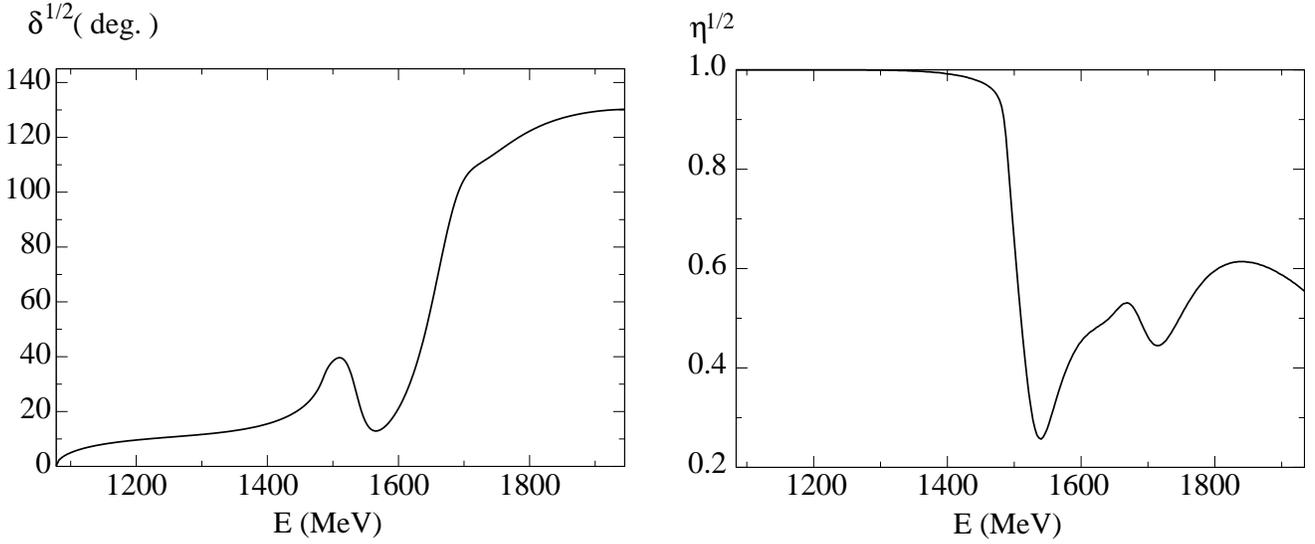

\includegraphics[scale=0.35]{Delta1h.eps}\quad\quad
\includegraphics[scale=0.35]{eta1h.eps}
\caption{Experimental phase shifts and inelasticity for the $\pi N$ interaction in isospin $1/2$.}\label{Exphase1h}
\end{figure}

\begin{figure}[ht!]
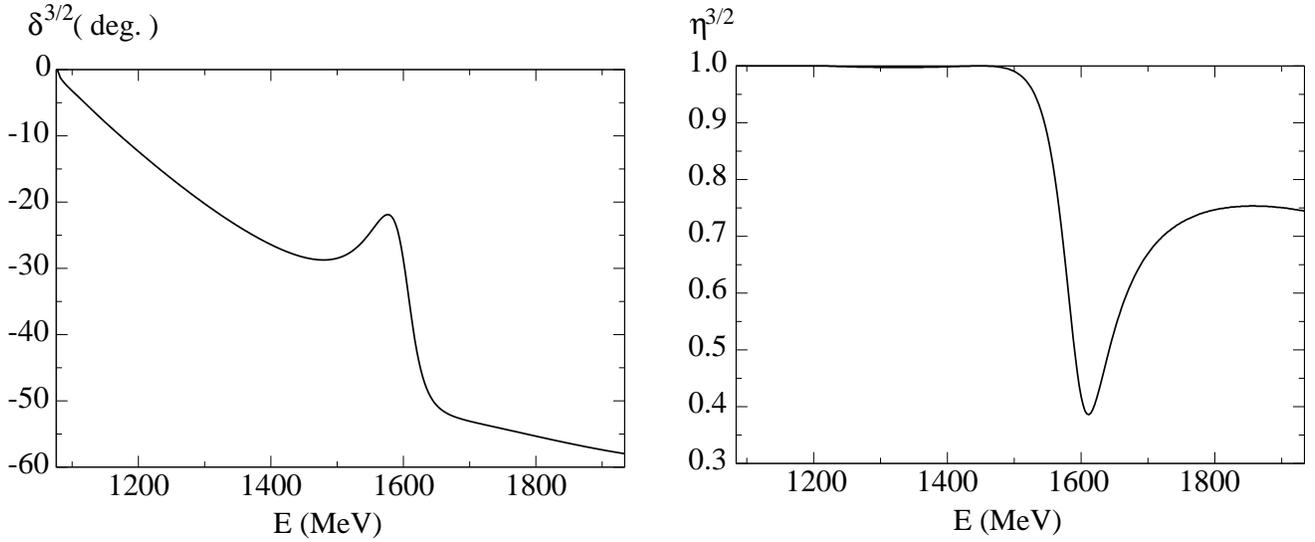

\includegraphics[scale=0.35]{Delta3h.eps}\quad\quad
\includegraphics[scale=0.35]{eta3h.eps}
\caption{Experimental phase shifts and inelasticity for the $\pi N$ interaction in isospin $3/2$.}\label{Exphase3h}
\end{figure}

\begin{figure}[ht!]
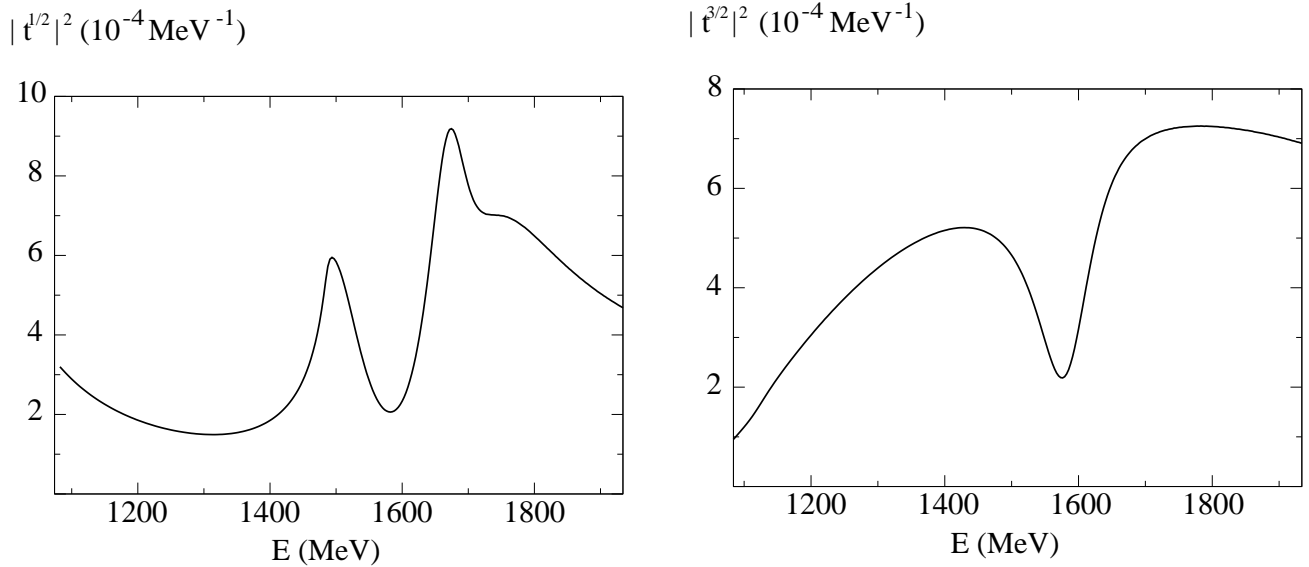

\includegraphics[scale=0.35]{t1h.eps}\quad\quad
\includegraphics[scale=0.35]{t3h.eps}
\caption{Experimental $t$-matrices for the $\pi N$ interaction in isospin $1/2$ and $3/2$.}\label{Ext} 
\end{figure}

The motivation of this work is to extend the calculations made in \cite{Khemchandani:2008rk} to higher energies by including the $N^*(1535)$ and $N^*(1650)$ in the input  $\pi N$ $t$-matrix and look for the other three-body isospin $1/2$ and $3/2$ states with $J^P=1/2^+$ in the $\pi\pi N$ system and coupled channels. In order to do this,  we use the experimental $L=0$ phase shifts ($\delta$) and inelasticities ($\eta$) \cite{Arndt} for the $\pi N$ system in isospin $1/2$ and $3/2$ configurations ( Fig. \ref{Exphase1h}, \ref{Exphase3h} )
and calculate from them the $\pi N$ amplitudes in the isospin base (Fig. \ref{Ext} ) using the relation
\begin{equation}
t^I=-\frac{4\pi E}{M} f^I,\quad\textrm{I=1/2,\,3/2}\label{texp}
\end{equation}
with
\begin{equation}
 f^I=\frac{\eta^I e^{2i\delta^I}-1}{2iq}
\end{equation}
where $\eta^I$ is the inelasticity, $\delta^I$ the phase shift, $M$ is the nucleon mass, $E$ is the $\pi N$ center 
of mass energy and  $q$ is the corresponding momentum.

\newpage
We require the input two-body  $t$-matrices in the charge base to solve the Faddeev equations  in our model. For this we use the relations
\begin{eqnarray}
t_{\pi^0 n\rightarrow\pi^0 n}&=&\frac{2}{3}t^{3/2}+\frac{1}{3}t^{1/2},\quad
t_{\pi^0 n\rightarrow\pi^- p}=\frac{\sqrt{2}}{3}t^{3/2}-\frac{\sqrt{2}}{3}t^{1/2},\nonumber\\
t_{\pi^- p\rightarrow\pi^- p}&=&\frac{1}{3}t^{3/2}+\frac{2}{3}t^{1/2},\quad
t_{\pi^- n\rightarrow\pi^- n}=t^{3/2},\nonumber\\
t_{\pi^+ n\rightarrow\pi^+ n}&=&t_{\pi^- p\rightarrow\pi^- p} ,\quad
t_{\pi^0 p\rightarrow\pi^0 p}=t_{\pi^0 n\rightarrow\pi^0 n},\\
t_{\pi^0 p\rightarrow\pi^+ n}&=&-t_{\pi^0 n\rightarrow\pi^- p}.\nonumber
\end{eqnarray}

Using these $\pi N$ $t$-matrices as input for Eqs. (\ref{Fa}), we can extend the model for the $\pi\pi N$ interaction of \cite{Khemchandani:2008rk} to higher energies where the invariant masses of the $\pi N$ subsystems can be varied around $1650$ MeV.

At this point we would like to discuss the cancellation between the off-shell part of the $t$-matrices in the Faddeev equations and the three-body forces, which justifies the use of the experimental amplitudes in our approach. These cancellations have been illustrated in all detail in the appendix of \cite{Khemchandani:2008rk} for the systems of two mesons and one baryon and in the appendix of \cite{MartinezTorres:2008gy} for the one vector-two pseudoscalar systems. The proof proceeded taking the potentials (tree level amplitudes) derived from the lowest order chiral Lagrangians. The extention of the proof made in the appendices of \cite{Khemchandani:2008rk,MartinezTorres:2008gy} to the corresponding one using $t$-matrices is straight forward, since the $t$-matrices would be generated by further iterations of the tree level amplitudes in the Faddeev diagrams, as done in \cite{Oller:1998zr,Inoue:2001ip},
where the off-shell part of the potential in these iterations is reabsorbed in constants of the on-shell potential. Hence the cancellations are guaranteed when iterations are done to obtain Faddeev diagrams in terms of $t$-matrices rather than potentials. This was discussed in \cite{Khemchandani:2008rk,MartinezTorres:2008gy}.

One could wonder if such a cancellation would also occur in those cases where the higher order terms of the Lagrangian would be necessary. Technically our assertion, that one can use only
the on-shell amplitudes, is rigorous as long as the amplitudes obtained with the lowest order chiral Lagrangian, upon unitarization, can reproduce the experimental data. This seems to be the case, for example, in S=-1 systems, for the energy range considered here. Indeed, calculations done in \cite{borasoy} using higher order terms of the Lagrangians show that the results obtained by using the lowest order Lagrangian fall well within the accepted uncertainties in the model. The situation is different for $S=0$, since, as mentioned in the introduction, the results of the calculations done with the lowest order chiral Lagrangian already fail beyond the total energy of 1600 MeV of the $\pi N$ system. Thus we could formally make no claims in this region about cancellations between the off-shell part of the $T_R$-matrices and the three-body forces. However, we also insist on the fact that the results cannot depend on the off-shell part  of the amplitudes, because these are unphysical. 

The finding of the exact cancellation of the off-shell part with the three-body forces is very useful because it implies that the Faddeev equations can be solved using only the physical information, that is the on-shell amplitudes. This feature certainly must sustain even when one goes beyond that realm where the lowest order chiral Lagrangian  reproduces the experimental data.
It would be interesting to study  cancellations similar to those found in \cite{Khemchandani:2008rk,MartinezTorres:2008gy} for the present case by using higher order Lagrangians but this is beyond the scope of the present work. 

It should be also said, when using higher order terms in the Lagrangians,  that although the elimination of the off-shell (unphysical) part is guaranteed, because the results cannot depend on unphysical amplitudes, it is not clear that the cancellation mentioned above would not leave some finite remanent part. It is also not guaranteed that, apart from three-body forces originating from the chiral Lagrangians, there are no other genuine three-body forces which would remain after necessary cancellations of off-shell terms.

However, let us make the following observation. We used a theory suited to the study of the $\pi\pi N$ system up to $\sqrt{s}\simeq 1750-1850$ MeV and concluded that one can study the system using only on-shell amplitudes, which one can get from experiments. Although proved within a certain theory, the conclusion that  ``one can solve Faddeev equations with experimental amplitudes'' is not linked to any model. With this in mind we make an ansatz that this conclusion should not be linked to the theory used to prove it and should be a characteristic of the dynamics of these systems for a wider range of energies than the one where we could establish a proof based on a particular theoretical framework. Afterall we will only extend our region of energies up to $\sqrt{s} \simeq$ 2200 MeV which is not too far from the energies at which the calculations were made earlier. Although certainly it is an ansatz at these higher energies, a posteriori, our assumption that one can rely solely upon the on-shell amplitudes in the Faddeev approach gets a strong support from the results that we obtain in the present work.

\subsection{Exploring the $\pi\pi N$ system}

We first study the $\pi\pi N$ system with total charge zero considering $\pi^0\pi^0 n$, $\pi^0\pi^- p$, $\pi^+\pi^- n$, $\pi^-\pi^+ n$ and $\pi^-\pi^0 p$ as coupled channels. We label them as particle $1$, $2$ and $3$ in the order in which they are written above. We calculate the three-body $T_R^{ij}$ matrices (Eqs. (\ref{Fa})) by using, for the $\pi N$ interaction:
(a) experimental amplitudes, i.e., Eq. (\ref{texp}) with the phase shifts and inelasticities shown in Figs. \ref{Exphase1h}, when the invariant mass of $\pi N$ system  is above its threshold (b) and the  $t$-matrix obtained from chiral Lagrangian \cite{Inoue:2001ip} for those $\pi N$ total energies which fall below the threshold.
For the $\pi \pi$ interaction we use the $t$-matrix obtained and studied thoroughly in \cite{Oller:1997ti}, where the dynamical generation of the $\sigma(600)$, $f_0 (980)$ and $a_0 (980)$ resonances was found and the theoretical results for physical observables coincided well with the experimental ones. We take proper symmetrized amplitudes into account wherever necessary, for instance, for the $\pi^0 \pi^0$ subsystem in the $\pi^0 \pi^0 N$ channel.

As in \cite{Khemchandani:2008rk}, we obtain the total $T^*_R$-matrix defined as $T^*_R \equiv $\smash{$\sum_{ij} (T_R^{ij} -t^ig^{ij}t^j)$} (see \cite{MartinezTorres:2007sr} for this definition). In order to identify the nature of the resulting states, we project the $T^*_R$-matrix on the isospin base. An appropriate base is the one where the states are classified by the total isospin of the three particles ``$I$'' and the total isospin of a subsystem ``$I_{sub}$''. We thus label the states in the isospin base as $\mid I, I_{sub} \rangle$. Obviously transitions between states with same total isospin but different isospin of a subsystem are possible. The peaks in the amplitudes are nevertheless seen more clearly for some particular isospin of the subsystem, indicating that the dominant structure of the state  found in the three-body system has a certain value of the total isospin and that of the isospin of a subsystem. We can thus write our $T^*_R$-matrix in the isospin base, in general, as 
$\langle I, I_{sub} \mid T^*_R (\sqrt{s}, \sqrt{s_{23}}) \mid I, I^\prime_{sub} \rangle$.

In order to be consistent with our previous work \cite{Khemchandani:2008rk}, we first check if we find an evidence  for the $N^* (1710)$. For this, we obtain the amplitude 
for total isospin of the three particles $I$=1/2 and for the isospin of the $\pi \pi$ subsystem, denoted by $I_{\pi\pi}$, being equal to zero, i.e.,
we calculate $\langle I = 1/2, I_{\pi\pi} = 0 | T_R (\sqrt{s}, \sqrt{s_{23}}) |  I = 1/2, I_{\pi\pi} = 0 \rangle$ (see Eqs. (30) of
\cite{Khemchandani:2008rk} for a more detailed definition of this isospin base).

We find exactly the same peak at $1704$ MeV in the squared 
amplitude as obtained in \cite{Khemchandani:2008rk}. In this way, we ensure that we reproduce our previous results  by using the experimental data for the $\pi N$ interaction above the $\pi N$ threshold. With this assurance, we now look for resonances in the higher energy region in same or other isospin configurations.

\begin{figure}[h!]
\centering
\includegraphics[scale=1]{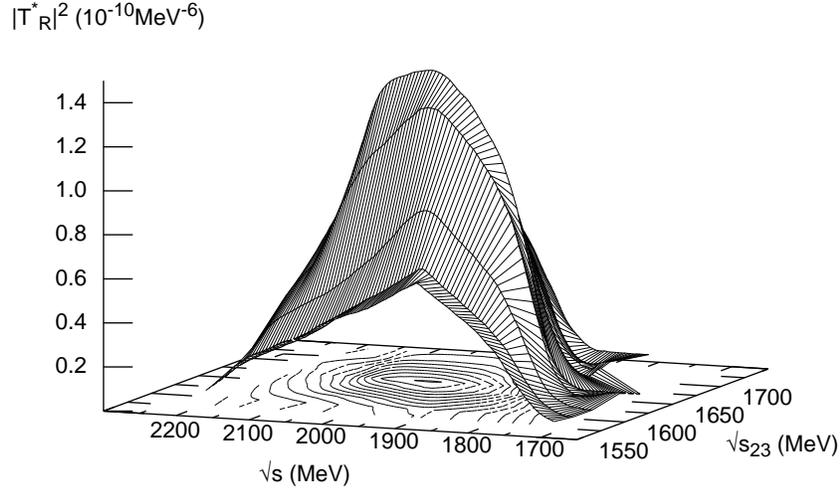}
\vspace{1cm}
\caption{The $N^*(2100)$ in the $\pi\pi N$ system with five coupled channels.}\label{N_5}
\end{figure}

We now study the $\pi\pi N$ amplitudes for the case in which the isospin of the subsystem of particle $2$ and $3$, i.e., pion nucleon (and its coupled channels) is $1/2$ in the initial as well as the final state.  To obtain this amplitude we write the $\pi \pi N$ states in the isospin base as

\begin{eqnarray}
\mid \pi^0\,\pi^0\,n\rangle&=&\mid 1, 0 \rangle \otimes \mid 1, 0 \rangle \otimes \mid 1/2, -1/2 \rangle\\
&=& \mid 1, 0 \rangle \otimes \left\{\sqrt{\frac{2}{3}}\mid I_{\pi N} = 3/2, I_{\pi N}^z = -1/2 \rangle + \sqrt{\frac{1}{3}}\mid I_{\pi N} = 1/2, I_{\pi N}^z = -1/2 \rangle \right\} \nonumber\\
&=&\sqrt{\frac{2}{5}}\mid I = 5/2, I_{\pi N} = 3/2 \rangle + \frac{1}{3}\sqrt{\frac{2}{5}} \mid I = 3/2, I_{\pi N} = 3/2 \rangle  - \frac{\sqrt{2}}{3}\mid I = 1/2,  I_{\pi N} = 3/2 \rangle +\nonumber \\ 
 &+& \frac{\sqrt{2}}{3}\mid I = 3/2,  I_{\pi N} = 1/2\rangle +
\frac{1}{3} \mid I = 1/2, I_{\pi N} = 1/2 \rangle.
\nonumber\label{iso1}
\end{eqnarray}
Similarly, by omitting the label $I$ and $I_{\pi N}$ to simplify,

\begin{eqnarray}
\mid \pi^+\, \pi^- \,n \rangle&=&-\sqrt{\frac{1}{10}}\mid 5/2, 3/2 \rangle - \sqrt{\frac{2}{5}}\mid 3/2, 3/2\rangle-
\sqrt{\frac{1}{2}}\mid 1/2, 3/2 \rangle  \\\label{iso2}
\mid \pi^-\,\pi^+\, n \rangle
&=&-\sqrt{\frac{1}{10}}\mid 5/2, 3/2\rangle 
 + {\frac{2}{3}} \sqrt{\frac{2}{5}} \mid 3/2, 3/2\rangle -
\frac{1}{3\sqrt{2}}\mid 1/2, 3/2 \rangle - \frac{\sqrt{2}}{3} \mid 3/2, 1/2 \rangle + {\frac{2}{3}}\mid 1/2, 1/2 \rangle
\nonumber\\
\mid \pi^-\,\pi^0\,p \rangle&=&\sqrt{\frac{1}{5}}\mid 5/2, 3/2 \rangle - \frac{4}{3\sqrt{5}}\mid 3/2, 3/2 \rangle +
\frac{1}{3}\mid 1/2, 3/2 \rangle - \frac{1}{3} \mid 3/2, 1/2 \rangle +
\frac{\sqrt{2}}{3}\mid 1/2, 1/2 \rangle
\nonumber\\
\mid \pi^0\,\pi^-\,p \rangle&=&\sqrt{\frac{1}{5}}\mid 5/2, 3/2 \rangle + \frac{1}{3\sqrt{5}}\mid 3/2, 3/2 \rangle -\frac{1}{3}\mid 1/2, 3/2 \rangle - \frac{2}{3}\mid 3/2, 1/2 \rangle -
\frac{\sqrt{2}}{3}\mid 1/2, 1/2 \rangle.\nonumber 
\end{eqnarray}

Inverting the above equations we get, for example,
\begin{eqnarray}
\mid 1/2, 1/2 \rangle&=&  \frac{1}{3}\biggr(  \mid \pi^0\,\pi^0\,n\rangle - \sqrt{2} \mid \pi^0\,\pi^-\,p \rangle +  \sqrt{2} \mid \pi^-\,\pi^0\,p \rangle + 2 \mid \pi^-\, \pi^+ \,n \rangle \biggr)\\ \nonumber
\mid 3/2, 1/2 \rangle&=& \frac{1}{3} \biggr( \sqrt{2}\mid \pi^0\,\pi^0\,n\rangle - 2  \mid \pi^0\, \pi^- \,p \rangle -\mid \pi^-\, \pi^0 \,p \rangle
- \sqrt{2}\mid \pi^-\, \pi^+ \,n \rangle  \biggr). \label{iso3}
\end{eqnarray}

In Fig. \ref{N_5} we show the squared $T^*_R$ amplitude for $I = 1/2$ and $I_{\pi N}$ =1/2 in the initial and the final state versus the total energy of the three-body system and the invariant mass of the meson-baryon subsystem formed by the second and third particle ($\pi N$).
A peak around an energy of $2100$ MeV with a width of $\sim$ 250 MeV  appears when $\sqrt{s_{23}}$  is close to 1670 MeV, thus having a $\pi N^*(1650)$ structure. The peak position and the width of this peak are compatible with the findings of various partial wave analyzes indicated by the PDG \cite{PDG} about the $N^*(2100)$, for which the peak position is found in the range 1855 - 2200 MeV and the width in the range of 69-360 MeV. Thus we identify this peak with the $N^*(2100)$.

Since this peak appears when $\sqrt{s_{23}}$ is close to the mass of the $N^*$ (1650) and has been obtained using as input that $\pi N$ $t$-matrix which contains the information on the $N^*(1650)$, we conclude that the inclusion of the $N^*(1650)$ in the $\pi N$ subsystem is essential to generate a resonance at $2100$ MeV.

In this former study, we do not find evidence for any resonance in the isospin $3/2$ configuration, but the situation is different when we introduce coupled channels, as we discuss below.
\begin{figure}
\centering
\includegraphics[scale=0.98]{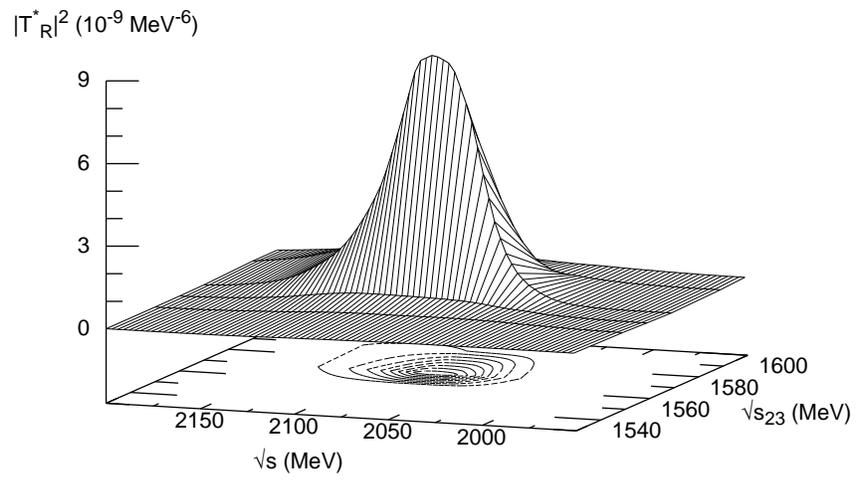}
\vspace{1cm}
\caption{The $N^*(2100)$ in the $\pi\pi N$ system including 14 coupled channels.}\label{N_14}
\end{figure}

\subsection{Inclusion of the $\pi K\Sigma$, $\pi K\Lambda$ and $\pi\eta N$ channels}
Next, we solve the Faddeev equations with fourteen coupled channels: $\pi^0\pi^0 n$, $\pi^0\pi^- p$, $\pi^0 K^+\Sigma^-$, $\pi^0 K^0\Sigma^0$, $\pi^0 K^0\Lambda$, $\pi^0\eta n$, $\pi^+\pi^- n$, $\pi^+ K^0\Sigma^-$, $\pi^-\pi^+ n$, $\pi^-\pi^0 p$, $\pi^- K^+ \Sigma^0$, $\pi^- K^0\Sigma^+$, $\pi^- K^+\Lambda$ and $\pi^-\eta p$. Again, we label them as particles $1$, $2$ and $3$ in the order in which they are written above. As there are no data for $K\Sigma \rightarrow K\Sigma$, $K\Lambda\rightarrow K\Lambda$, etc., we use the model of \cite{Inoue:2001ip} to calculate the corresponding amplitudes. The $\pi N$ interaction below threshold is determined using the same model as for $K\Sigma$ and $K\Lambda$ and above the threshold we use the experimental results.

We continue to study those amplitudes where the isospin of the subsystem of particle $2$ and $3$, i.e., pion nucleon (and its coupled channels), is $1/2$ in the initial as well as the final state. 
In Fig. \ref{N_14} we show the $\pi \pi  N$ amplitude  for total isospin $I=1/2$ for such a case. 

As shown in Fig.\ref{N_14} we obtain a peak at an energy of $2080$ MeV with a width of $54$ MeV for a $\sqrt{s_{23}}$ near $1570$ MeV, which we identify with the $N^*(2100)$ listed in the PDG \cite{PDG}. Comparison of the Figs. \ref{N_5} and \ref{N_14} shows that the inclusion of the $\pi K\Sigma$, $\pi K\Lambda$ and $\pi\eta N$ channels makes the resonance more pronounced (by an order of magnitude in the squared $T^*_R$-matrix) and much narrower. These changes in the results can be easily understood with respect to the previous ones obtained with only five coupled channels by noticing that now the wave function of the resonance contains extra components
which have smaller phase space in the decay of the resonance. At the same time, the $\pi\pi N$ component becomes smaller due to the normalization of the wave function and, hence, the decay into $\pi\pi N$ is also reduced.

\begin{figure}[ht!]
\centering
\includegraphics[scale=0.98]{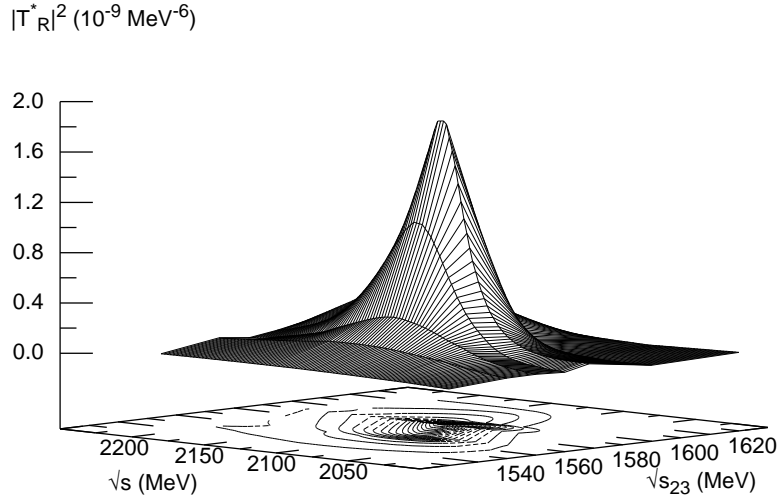}
\vspace{1cm}
\caption{The $\Delta(1910)$ in the $\pi K\Lambda$ system including 14 coupled channels.}\label{D_14}
\end{figure}

On exploring other isospin configurations we find a peak in the $\pi K \Lambda$ amplitudes with total isospin $I$=3/2 and with isospin $I_{K\Lambda}$ = 1/2 in the initial as well as in the final state. In order to get this amplitude we have used the relation
\begin{equation}
\mid I=3/2, I_{K\Lambda}=1/2 \rangle = \frac{1}{\sqrt{3}} \big(\sqrt{2} \mid \pi^0 K^0 \Lambda \rangle + \mid \pi^- K^+ \Lambda \rangle \big), \label{iso4}
\end{equation}
which has been obtained by writing the $\pi K\Lambda$ states in isospin base analogously to Eqs. (\ref{iso1}, \ref{iso2}). In Fig. \ref{D_14} we show the squared $\langle I=3/2, I_{K \Lambda} = 1/2 \mid T^*_R (\sqrt{s}, \sqrt{s_{23}}) \mid I=3/2, I_{K\Lambda}=1/2 \rangle$ amplitude . A peak is found at a total energy of $\sim$ 2126 MeV with $\sim$ 42 MeV of width. In this case, the invariant mass $\sqrt{s_{23}}$, at which the peak appears, is around 1590 MeV. This peak can be identified with the $\Delta (1910)$ listed in  \cite{PDG}, whose position, given by different partial wave analyzes, ranges up to 2070 MeV and the width varies from 190-500 MeV.

Thus the introduction of the $\pi K\Sigma$, $\pi K\Lambda$ and $\pi\eta N$ channels, together with  the inclusion of the $N^*(1650)$ in the $\pi N$ $t$-matrix,  is important to get this resonance.
One should note that we get smaller widths than the experimental ones. The $\pi N$ decay channels are not considered in our approach and they should contribute to increase the widths. Note that this can be done even with a small $\pi N$ component, as implicitly assumed here, since there is more phase space for decay into the $\pi N$ channel (see \cite{MartinezTorres:2007sr} for more discussion).

We do not find any evidence of the $\Delta(1750)$, which could indicate a different structure for this state that the one studied in this work.

\section{Exploring the $Nf_0$ and $Na_0$ systems by taking $N\pi\pi$, $NK\bar{K}$ and $N\pi\eta$ as coupled channels}
Until now, we have investigated possible resonant states in the $\pi\pi N$ system and its coupled channels which have been obtained by adding a pion to pseudoscalar-baryon systems which couple strongly in $J^\pi=1/2^-$ and isospin $1/2$ configuration, i.e., $\pi N$, $K\Sigma$, $K\Lambda$ and $\eta N$. The invariant mass of this pseudoscalar-baryon subsystem has been varied around that of the $N^*(1535)$ and $N^*(1650)$, hence, treating the three-body system as a $\pi N^*$ system with $1500<M_{N^*}<1760$ MeV, although within the three-body Faddeev equations.  There are other configurations of this three-body system, like $Na_0(980)$ and $N f_0(980) $, which we have not discussed so far.

In order to study such a system, we must take $N K\bar{K}$, $N\pi\pi$ and  $N\pi\eta$ as coupled channels, such that the $\pi\pi$ and $K\bar{K}$ subsystem dynamically generate the $f_0(980)$ and the $\pi\eta$ subsystem along with $K\bar{K}$ generates the $a_0(980)$ resonance. In this way, we can study the $N f_0(980)$ and $Na_0(980)$ systems simultaneously.
Concretely, we take the following coupled channels into account: $n \pi^0 \pi^0$, $p \pi^0\pi^-$, $n \pi^0 \eta$, $n \pi^+ \pi^-$, $n \pi^- \pi^+$, $p \pi^- \pi^0$, $p \pi^- \eta$, $n K^+ K^-$, $n K^0 \bar{K}^0$, $p K^0 K^-$. We label the particles as $1$, $2$, $3$ in the order in which they are written above. This means that the subsystem of particles $2$ and $3$ consists of two pseudoscalar mesons whose invariant mass, $\sqrt{s_{23}}$, is varied around 980 MeV.
With these channels we solve the Eqs. (\ref{Fa}) in the same formalism which we have explained in the previous sections. In this case we find that
the $N K \bar{K}$ amplitude is bigger in magnitude as compared to those of the other coupled channels.
We thus make isospin combinations of the $N K \bar{K}$ channels, similarly to Eqs. (\ref{iso1}, \ref{iso2}) and obtain the amplitude for total isospin $I$ = 1/2 and the isospin of the $K \bar{K}$ system, $I_{K \bar{K}}$, equal to 0 or 1. 

In the case of  total isospin of the $N K \bar{K}$ system equal to $1/2$ with the isospin of the $K \bar{K}$ subsystem  equal to one the amplitude
\begin{equation}
\langle  I = 1/2, I_{K \bar{K}}=1 \mid T^*_R (\sqrt{s},\sqrt{s_{23}}) \mid I = 1/2, I_{K \bar{K}}=1 \rangle,
\end{equation}
shows a peak around $2080$ MeV , with a width of 51 MeV (which we do not show here), which we relate as the $Na_0(980)$ partner of the peaks shown in Figs. \ref{N_5} and \ref{N_14}. Thus the peak corresponding to the $N^*(2100)$ has been seen in $\pi \pi N$ system as well as in the $N K \bar{K}$ system.

\begin{figure}[h!]
\centering
\includegraphics[scale=1.2]{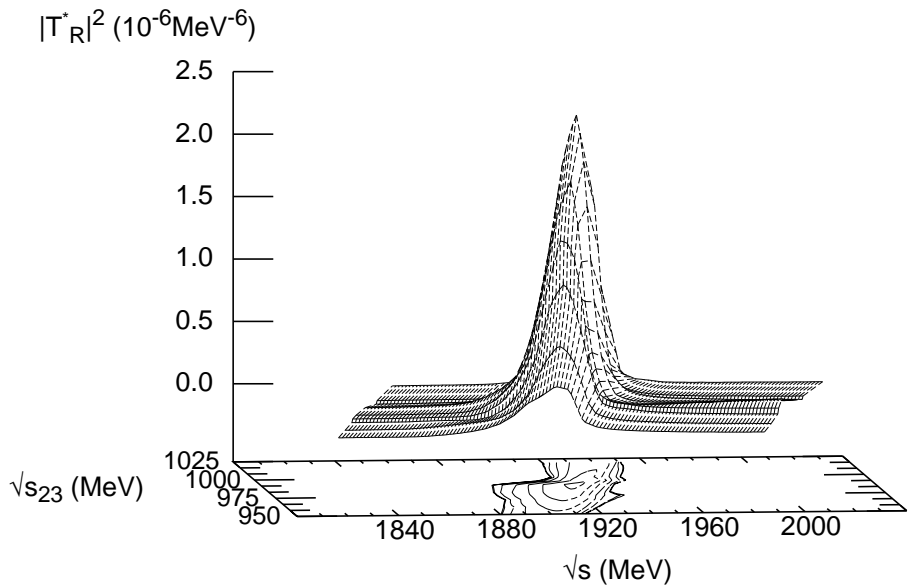}
\vspace{1cm}
\caption{A possible $N^*(1910)$ in the $N K\bar{K}$ channels.}\label{N_1910}
\end{figure}

Interestingly, along with this $N^*(2100)$ state, we find another peak with even larger magnitude of the squared three-body amplitude at $\sqrt{s}=1924$ MeV with a width of $20$ MeV. We show this peak in Fig. \ref{N_1910} for the $NK\bar{K}$ channel. 

This state is about $7$ MeV below the $NK\bar{K}$ threshold (assuming an average mass for the kaons of $496$ MeV and $939$ MeV for the nucleon). Therefore, this result indicates that the $Na_0(980)$ system gets bound at around $1920$ MeV. This possibility has been already suggested by the authors in \cite{Jido:2008kp}, in which they study the $NK\bar{K}$ channel using effective two-body potentials to describe the $\bar{K}N$, $\bar{K}K$, $K N$ interactions. They find that the $NK\bar{K}$ system can get bound while the $K\bar{K}$ subsystem acts like the $a_0$.  Our result is, thus, in agreement with the suggestions in \cite{Jido:2008kp}. Interestingly, the existence of a $1/2^+$ $N^*$ resonance around $1935$ MeV has also been proposed earlier \cite{Mart:1999ed} on the basis of a study of the data on the $\gamma p\rightarrow K^+\Lambda$ reaction in an isobar model, although other theories \cite{scholten1} which include explicitly resonances up to $1855$ MeV can reproduce these data (though further work along these lines to include higher mass resonances is under way \cite{scholten2}). 

Since this peak found at $1920$ MeV is below the three-body threshold and, in the two-body problem, the poles for the $f_0(980)$ and $a_0(980)$ appear below the $K\bar{K}$ threshold, the three particles in the system have associated complex momenta in the momentum representation. To avoid the use of unphysical complex momenta in the three-body system, which will lead to imaginary energies in the real plane, we give a minimum value, around 50 MeV, to the momentum of the particles. We have check the sensitivity of our results to the mentioned choice by changing the minimum momentum from 50 MeV to 100 MeV and we find the peak and width to remain almost unchanged.

We also made the total isospin $1/2$ combination of the $N K \bar{K}$ system by considering the $K \bar{K}$ subsystem in isospin 0 in the initial and the final state, i.e., considering the $Nf_0(980)$ component of the $N K \bar{K}$ channel. In this case too, just as in the amplitude for $N a_0(980)$, we find a peak around $1923$ MeV with a width of $30$ MeV and another one around $2052$ MeV with a width of $60$ MeV. The magnitude of the $Nf_0(980)$ amplitude around $1920$ MeV is very similar to the one of the $Na_0(980)$ amplitude (i.e., the one shown in Fig. \ref{N_1910}), but the magnitude of $Nf_0(980)$ amplitude around 2050 MeV is bigger than the magnitude of the  $Na_0(980)$ amplitude. 

From the whole study we would conclude that there are two $N^*$'s with $J^\pi=1/2^+$ in the energy region $1800<\sqrt{s}<2200$ MeV.

The peaks obtained in this approach are very neat and we associate them to physical resonance states. In the two body scattering it is customary to look for poles in the second Riemann sheet to associate them to resonances. The issue of poles for the three body problem within our formalism was addressed in the section VI of \cite{MartinezTorres:2008gy}. The difficulty to work with two complex variables, $\sqrt{s}$ and $\sqrt{s}_{23}$ which induce complex three momenta needed in the evaluation of integrals are obvious. Yet, in \cite{MartinezTorres:2008gy} an approximate method was devised, which is suitable for the present context too, for the case when a subsystem of two particles can be treated as a resonance. Therefore  the three-body system can be interpreted as a system of a particle and a resonance. This was the case for $\phi f_0(980)$ in \cite{MartinezTorres:2008gy} and the systems where resonances have been found in the present work can be treated similarly, for example, the $N K\bar{K}$ system where a resonance around 1920 MeV is found can be treated as a $N a_0 (980)$ system. In such cases, the problem can be reduced to a two-body scattering and usual poles can be identified in the complex energy ($\sqrt{s}$) plane. In as much as the resonances found here follow an approximate Breit-Wigner shape, as in \cite{MartinezTorres:2008gy}, the poles in the second Riemann sheet are guaranteed, as was shown in \cite{MartinezTorres:2008gy}. The amplitudes obtained in the present case  indeed follow approximate Breit-Wigner distributions, and as in 
\cite{MartinezTorres:2008gy}, they should have the corresponding poles in the $\sqrt{s}$ complex plane.

\section{Conclusions}
To summarize, we have extended our previous study of the $\pi\pi N$ system and coupled channels \cite{Khemchandani:2008rk}, where the generation of the $N^*(1710)$ was found, to higher energies. In this work the new input is the experimental data on the $\pi N$ interaction where the information on excitation of both the $N^*(1535)$ and the $N^*(1650)$ is present, the latter of which was absent in our previous work \cite{Khemchandani:2008rk}. Here, apart from confirming the $N^*(1710)$, we find evidence for the other $1/2^+$ $N^*$, i.e., the $N^*(2100)$, and also  for the $1/2^+$ $\Delta (1910)$ resonance. The findings reported here indicate that the inclusion of the $N^*(1650)$ in the interaction of the $\pi N$ subsystem is essential to generate these higher mass $1/2^+$ resonances. We have first made a search taking only the $\pi\pi N$ channels where a resonance having the properties of $N^*(2100)$ was found. Later we included the $\pi K\Sigma$, $\pi K\Lambda$ and $\pi \eta \Sigma$ channels where the same resonance is produced but with larger magnitude and narrower width, indicating the addition of more channels to which the resonance couples strongly. No isospin $3/2$ resonances is found in the study of the $\pi\pi N$ channels alone. However, the $\Delta(1910)$ is found on inclusion of the  $\pi K\Sigma$, $\pi K\Lambda$ and $\pi \eta \Sigma$ channels. Further, we have investigated the $NK\bar{K}$, $N\pi\pi$ and $N\pi\eta$ channels where the $K\bar{K}-\pi\pi$ subsystem rearranges itself as a $f_0(980)$ resonance, while $K\bar{K}-\pi\eta$ acts like the $a_0(980)$. We obtain a new peak at $\sim 1924$ MeV, apart from the one corresponding to the $N^*(2100)$, with a strong coupling to $Na_0(980)$ and $Nf_0(980)$. Finally, we conclude this work by stating that the study of three-body systems, for the cases where a complete theoretical two-body input is not available, is also possible in our formalism  using on shell experimental amplitudes.

\section*{Acknowledgments}  
We would like to thank Juan Nieves for discussions and for providing us the
experimental $\pi N$ amplitudes.
This work is partly supported by DGICYT contract number
FIS2006-03438 and  the JSPS-CSIC
collaboration agreement no. 2005JP0002, and Grant for Scientific
Research of JSPS No.188661.
One of the authors (A. M. T) is supported by a FPU grant of the Ministerio de Ciencia y Tecnolog\'ia.
K. P. Khemchandani thanks the support by the
\textit{Funda\c{c}$\tilde{a}$o para a Ci$\hat{e}$ncia e a Tecnologia of the Minist\'erio da Ci$\hat{e}$ncia, Tecnologia e Ensino Superior} of Portugal (SFRH/BPD/40309/2007). This research is  part of
the EU Integrated Infrastructure Initiative  Hadron Physics Project
under  contract number RII3-CT-2004-506078.

\end{document}